\documentclass[twocolumn,prb,aps,superscriptaddress,showpacs]{revtex4}
\usepackage{graphicx}
\begin{document}
\title {Density-induced reorientation of the stripe at half-filled high Landau levels}
\author{Shi-Jie Yang}
\affiliation{Department of Physics, Beijing Normal University,
Beijing 100875, China}
\affiliation{Center for Advanced Study,
Tsinghua University, Beijing 100084, China}
\author{Yue Yu}
\affiliation{Institute of Theoretical Physics,
Chinese Academy of Sciences, P.O. Box 2735, Beijing 100080, China}

\begin{abstract}
The effect of a unidirectional periodic potential on the
orientation of the stripe state is studied for the two-dimensional
electron system at half-filled high Landau levels. By considering
a quantum well with two electric subbands, it is found that the
stripe is parallel to the external potential for weak modulation
and is orthogonal for strong modulation. In the intermediate
range, the orientation of the stripe changes from orthogonal to
parallel as the electron density is increased. This result
explains the recent experiment performed by J. Zhu {\it et al}
that the anisotropy axis at half-filled high Landau levels rotates
by $90^0$ by increasing the electron density. It also supports the
suggestion that the stripes is pinned by the native surface
morphology at the interface of the heterojunction.

\end{abstract}
\maketitle
%\pagenumbering {arabic}

Two-dimensional electron systems (2DES) of the high mobility
samples in a strong magnetic field exhibit a rich variety of
physical phenomena associated with the Coulomb interactions
between electrons. In the lowest Landau level (LL), the electrons
condense into an incompressible quantum liquid at certain rational
filling factors and leads to the famous fractional quantum Hall
effect (FQHE)\cite{Tsui,Laughlin,book1}. Recently, a new kind of
many-body correlated phase which shows large anisotropy in the
longitudinal resistance in the magneto-transport experiments, was
revealed at half-filled high Landau levels ($\nu=9/2,11/2,\cdots$)
\cite{Lilly1,Du1}. The origin of this strange anisotropy is widely
viewed as the formation of a unidirectional charge density wave
(UCDW) or stripe state around these filling
factors\cite{Phillips,Jungwirth}. Even before the experimental
discoveries, the UCDW was predicted theoretically by M.M. Fogler
{\it et al} and R. Moessner{\it et al} based on the Hartree-Fock
(HF) discussions, where the 2D electron gas spontaneously breaks
the translational symmetry\cite{Fogler,Moessner}. When the
external field is tilted away from the sample normal, it shows
that the easy transport orients orthogonal to the in-plane
magnetic field\cite{Lilly2,Du2}. Theoretical computations beyond
the HF approximation are also consistent reasonably with the
experiment results \cite{Rezayi,MacDonald,Shibata}. Some
researchers proposed the existence of liquid crystalline states
with stripe ordering and broken rotational
symmetry\cite{Fradkin,Fertig,MacDonald}.

Nevertheless, the preferred orientation of the stripes in a
perpendicular magnetic field remains puzzling\cite{Rosenow}. For
2D electron systems in GaAs/AlGaAs heterostructures grown on
$<110>$-oriented GaAs substrates, the hard transport direction is
parallel to the $<1\bar 10>$ crystallographic direction while the
easy direction is parallel to $<110>$. It is hard to believe that
the crystal structure affects the orientation of the stripe. In
the experiments the magnetic field is a few teslas, so the wave
function of each electron in the third LL is spread over more than
several hundred angstroms. The details of the crystal lattice
structure will be averaged out. Fil proposed that the
piezoelectric effect may play a role in determining the
orientation of the stripes\cite{Fil}. In a recent experiment, R.L.
Willett {\it et al} examined the surface morphology of high
mobility heterostructures and found the transport is consistent
with that in samples having artificially induced 1D charge
modulations\cite{Willett}. The native lines are orthogonal to the
stripes, which at first glance, is somehow contrast to intuitions.
Several authors have studied this new effect with a periodic
external potential and their results consistent with the
experiment\cite{Aoyama,Yoshioka}.

Recently, J. Zhu {\it et al} observed a density-induced
interchange of anisotropy axes at half-filled high LLs\cite{Zhu}.
They employed a tunable density heterostructure insulated gate
field effect transistor to access a wide density regime and found
that as the density of the 2DES is raised above $2.9\times 10^{11}
cm^{-2} $, the easy axis rotates from the $<110>$ direction to the
$<1\bar 10>$ direction. Their result provides another way to
demonstrate the pinning mechanism of the stripe phase which we
will discuss in the this work. We will show when take account into
two electric subbands and introduce a unidirectional periodic
potential to the electron system, these stripes align either
parallel or perpendicular to the external potential. The stripe is
parallel to the periodic potential for weak modulation whereas it
perpendicular to the potential for strong modulation. For
intermediate modulation, the stripes experience a rotation of
$90^0$ from parallel or orthogonal to the external potential as
the density increases. Our discussion supports the suggestion that
the orientation of the stripe phase is pinned by the native
symmetry breaking potential at the GaAs/AlGaAs interface.

In order to deal with the problem analytically, we assume that the
electron gas is confined in a plane by an harmonic potential with
the characteristic frequency $\Omega$. Before going to the
details, we argue that this choice of the confining potential may
quantitatively correct for the problem we will deal with in this
work despite the fact that the realistic confining potential in
the sample is essentially a finite square well. The harmonic well
is very different from the square well in their excited spectra,
for the harmonic spectrum is equal gapped while that of the square
well is not. However, in our work, there will be only two energy
levels of the confining potential to be involved and thus the
unequal energy gaps between different adjacent energy level will
not be concerned. Therefore, one can variationally adjust the
harmonic frequency such that the harmonic energy gap equals the
gap between those two given levels in a realistic square well. In
this sense, the harmonic potential may be a good approximation to
a realistic potential to give a quantitative description. Such a
harmonic potential has been chosen to deal with many quantum Hall
systems to replace the realistic potential which is either
triangular\cite{cha} or square\cite{Phillips,Jungwirth,yu}. It was
also used to discuss the giant magneto-resistance induced by a
parallel magnetic field \cite{das}.

For a perpendicular magnetic field applied to the system, ${\bf
B}=B\hat{z}$, there are two electric subbands mixing with the
Landau levels, with frequencies $\omega_{+}=\Omega$ and
$\omega_{-}=\omega_c$, respectively. The corresponding eigen
wavefunctions are\cite{Yang}
\begin{eqnarray}
\phi_{m}^{\omega_{+}}&=&N_{m}^{+}e^{-z^2/2l_{+}^2}H_{m}(z/l_{+}) \nonumber \\
\phi_{n}^{\omega_{-}}&=&N_{n}^{-}e^{-(x-X)^2/2l_{-}^2}H_{n}((x-X)/l_{-}),
\end{eqnarray}
where $H_{n}(x)$ are the Hermitian polynomials and
$N_n^{\pm}={1\over \sqrt{2^n n!\sqrt{\pi}l_{\pm}}}$ are the normalization
coefficients. $l^2_{\pm}=\hbar/m_b \omega_{\pm}$. $X$ is an integer
multiple of $2\pi l_{-}^2/L_y$. The combined single particle wavefunction
is
\begin{equation}
\Phi_{mn}=\frac{1}{\sqrt{L_y}}e^{iXy/l^2}\phi_{m}^{\omega_{+}}
          \times \phi_{n}^{\omega_{-}}.
\label{one}
\end{equation}
Then the energy levels of single particle states are described by a
set of two indices $(mn)$ with index $m$ indicating the electric
subbands and $n$ the Landau levels.

Fig.1 schematically depicts the energy levels of the 2DES. Given a
filling factor, {\it e.g.}, $\nu=9/2$, increasing the electron
density means increasing the strength of the magnetic field $B$.
Hence there appear a series of energy level crossings since the
electric subbands rarely change whereas the cyclotron frequency
$\omega_c$ increases with the magnetic field. The Fermi energy is
indicated by the thick dashed line. Since the single particle
state at the Fermi surface is changed from $(10)$ on the left to
$(01)$ on the right of the crossing point as the density
increases, one may expect a phase transition at the level crossing
as in most circumstances\cite{Pan,Sarma}. We will show that the
UCDW indeed changes from orthogonal to parallel to the periodic
potential for intermediate modulations as the electron density
(hence the magnetic field) increases.

To include the surface morphology at the interface of the
heterostructure, we consider the effect of a periodic potential of
wave vector $\vec{Q}_p$ and strength $V_0$ on the stripes of
fundamental wave vector $\vec{Q}_s$. In realistic samples of
GaAs/AlGaAs heterostructure, the wave vector of potential
modulation $\vec{Q}_p$ should be in the $<110>$
direction\cite{Willett,Cooper,Orme}. Two configurations of
$\vec{Q}_p$ and $\vec{Q}_s$ are considered in our work:
$\vec{Q}_p\parallel \vec{Q}_s$ or $\vec{Q}_p\perp \vec{Q}_s$. In
the orthogonal orientation, the main deformation of the stripe
caused by the periodic potential is modulation of the width of the
stripes. In the parallel orientation, the main deformation of the
stripe is displacement of the stripes. Both deformations lower the
cohesive energy of the stripe state\cite{Yoshioka}.

The Hamiltonian consists of a Coulomb interaction part $H_0$ and an
interaction with the external potential part $H_1$: $H=H_0+H_1$. Here
\begin{equation}
H_0=\frac{1}{2L_xL_y}\sum_{\vec{q}}v(\vec{q})\rho (\vec{q})\rho(-\vec{q}),
\label{two}
\end{equation}
where $v(\vec{q})=2\pi e^2/\kappa_0 q$.
\begin{equation}
H_1=\frac{1}{2}V_0 \sum_{\vec{q}=\pm\vec{Q}_p}\rho(\vec{q})
     exp(i\vec{q}\cdot \vec{r_0}),
\end{equation}
where $\vec{r}_0$ is the origin of the potential. $\rho(\vec{q})$
is the electron density operator projected onto the upper LL. It
can be written as
\begin{equation}
\rho(\vec{q})=\sum_{X}F(\vec{q})e^{-iq_x X}c^{\dagger}_{X_{+}}c_{X_{-}},
\end{equation}
where $X_{\pm}=X\pm q_y l^2/2$. $F(\vec{q})$ is computed by the
state (\ref{one}), which is given by
\begin{equation}
F(\vec{q})=e^{-q_z^2l_+^2/4-q^2_{\Vert}l_{-}^2/4}L_m(q^2_zl_+^2/2)
     L_n(q_{\Vert}^2l_{-}^2/2),
\label{six}
\end{equation}
where $L_n(x)$ is the Laguerre polynomial. $\vec{q}_{\Vert}$ is the
momentum in the 2DES plane.

By using the standard manipulation for the Hartree-Fock decoupling
of Hamiltonian (\ref{two}), we get
\begin{equation}
H_0^{HF}=\frac{1}{2}\sum_{\vec{q}}u_{HF}(\vec{q})\Delta(-\vec{q})
         \sum_{X}e^{-iq_xX}c^\dagger_{X_+}c_{X_-},
\end{equation}
The effective potential $u_{HF}(\vec{q})$ is explicitly written
as a sum of a Hartree term (in units of $e^2/\kappa_0 l$)
\begin{equation}
u_H(\vec{q})=\int\frac{dq_z}{\pi l}\frac{1}{q^2_\Vert+q^2_z}[F(\vec{q})]^2,
\end{equation}
and an exchange term
\begin{equation}
u_{ex}(\vec{q})=-2\pi l^2\int \frac{d\vec{p}}{(2\pi)^2}
                u_H(\vec{p})e^{i\vec{p}\times \vec{q}l^2}.
\end{equation}
Allowing the charge density wave (CDW) by introducing order parameters
\begin{equation}
\Delta(\vec{Q})=\frac{2\pi l^2}{L_xL_y}\sum_{X}e^{-iQ_xX}
                <c^\dagger_{X_+}c_{X_-}>,
\end{equation}
the cohesive energy of the electrons in the topmost LL can be obtained as
\begin{equation}
E_{coh}=\frac{1}{2\nu_N}\sum_{\vec{Q}\neq 0}u_{HF}(\vec{Q})
        |\Delta(-\vec{Q})|^2,
\end{equation}
where $\nu_N=1/2$ is the filling factor at the topmost Landau
level.

We carry out a HF computation on a rectangular lattice with the
wave vectors of the order parameters as $\vec{Q}=(jQ_x^0,kQ_y^0)$,
where $j$ and $k$ are integers. Following the procedure in
Refs.[\onlinecite{Yoshioka,Yoshioka2}], when
\begin{equation}
NQ_x^0Q_y^0l^2=2M\pi
\end{equation}
with $N$ and $M$ being integers, the Landau level splits into $N$
Hofstadter bands. The CDW state is recognized as the stripe phase
when the order parameters with $\vec{Q}=\pm \vec{Q}_s$ are
dominant. When $N=6$ and $M=1$ the stripe state has the lowest
energy. In ref.[\onlinecite{Yoshioka}], it considered several
cases for $Q_p=Q_s/k$, with $k=2,3,4,5,6$. As an example in our
work, the wave vector of the external potential is typically
chosen to be $Q_p=Q_s/3$. We consider two configurations in which
the $\vec{Q}_p$ and $\vec{Q}_s$ are either parallel or orthogonal
to each other, respectively. In the parallel orientation case, the
main deformation is the displacement of the stripe, which is
called the "frequency modulation". In the orthogonal orientation
case, the main deformation is the modulation of the stripe width,
which is called the "amplitude modulation". In the latter case,
there is a periodic density modulation along the stripes with wave
vector $\vec{Q}=\vec{Q}_p$ for weaker modulation. When the
modulation becomes stronger ($V_0/\hbar\Omega\gtrsim 0.1$), the
stripe looks like breaking up at the ridges of the external
potential and degenerates into a rectangular CDW state.

Fig.2 shows the dependence of the cohesive energy of the electrons
in the third LL on the modulation strength of the external
potential $V_0$. The parallel orientation state and orthogonal
orientation state for $(mn)$ energy levels are denoted by
"para$(mn)$" or "orth$(mn)$", respectively. When the electron
density (or $\omega_c/\Omega$) raises, the Fermi level changes
from the $(10)$ state to the $(01)$ state. Fig.2(a) is for
$\omega_c/\Omega=0.2941$ and (b) is for $\omega_c/\Omega=0.8824$.
Both figures show that the parallel orientation is slightly lower
in energy than the orthogonal orientation for small modulation
strength ($V_0/\hbar\Omega\lesssim 0.04$) whereas the orthogonal
orientation dominant for large modulation strength
($V_0/\hbar\Omega\gtrsim 0.06$). Previous studies claimed that the
orientation of the stripe is always perpendicular to the periodic
potential\cite{Aoyama,Yoshioka}. The difference may be resulted
from that their calculations did not count in the width of quantum
well. We note that the two-subband is not the key element to
whether the stripes are parallel or orthogonal to the potential.
The orientation of the stripes is mainly dependent on the relative
strength of the external modulation $V_0$ with respective to the
characteristic frequency $\Omega$. However, in the single band
model, there is no transition of orientation of the stripes as the
density varies. In the two-subband levels, since the matrix
elements in formula (\ref{six}) are dependent on the single
particle states (\ref{one}), the Hartree-Fock potentials are
different at the two sides of the crossing, which the orientation
transition underlie.

Fig.3 shows the anisotropy energy $E_a$ versus $\omega_c/\Omega$
or electron density (in arbitrary units). $E_a$ is the energy
difference between the parallel orientation and the orthogonal
orientation. We depict three curves for three typical values of
$V_0/\hbar\Omega$. We find that $E_a$ is definitely negative for
$V_0/\hbar\Omega\lesssim 0.04$, indicating that the parallel
orientation is favored. $E_a$ is definitely positive for
$V_0/\hbar\Omega\gtrsim 0.06$, indicating the orthogonal
orientation is favored. For the curve $V_0/\hbar\Omega=0.05$,
$E_a$ changes from positive to negative as $\omega_c/\Omega$ (or
electron density) increases, implying a phase transition from the
orthogonal orientation to the parallel orientation. This result
coincides with the recently experimental observation by J. Zhu
{\it et al} that the anisotropy axes at half-filled high Landau
levels in the two-dimensional electron system rotates by $90^0$ by
increasing the electron density\cite{Zhu}. It should be noted that
in previous studied samples with electron densities inside the
transition region of Ref.[\onlinecite{Zhu}], the easy direction is
always parallel to the $<110>$
direction\cite{Lilly2,Du2,Willett,Cooper}. This variety of
experimental results may originate from the sensitive dependence
of the stripe orientation on the roughness of the surface
morphology at the interface of the heterostructure. In our
calculations, the parallel phase exists only for rather weak
modulation ($V_0/\hbar\Omega\lesssim 0.04$). Specifically,
anisotropy axis rotation takes place only in a limited range of
modulation strength and width of quantum well ($0.04 \lesssim
V_0/\hbar\Omega \lesssim 0.06$). Beyond this range, no
reorientation transition can be observed.

In fig.1 there are more energy crossings as $\omega_c/\Omega$ (or
electron density) further increases or decreases. For lower
density, the single particle state will be the $(02)$ state. The
stripe will experience an additional orientation interchange,
which may be observed in experiment by sweeping a larger electron
density regime. We emphasize that the complex transport behavior
takes place only in a two-subband quantum well at half-filled
Landau levels. Given a filling factor, increasing electron density
means increasing the magnetic field, which leads to energy level
crossings. Ref.[\onlinecite{Zhu}] suggested that the squeeze of
the electron wave function and press harder against the interface
may be the origin of reorientation of the stripes. We have checked
in our computations that no such transition can take place in a
single band model. Our result provides a support to the
explanation that the pinning mechanism is the native surface
morphology at the interface of the GaAs/AlGaAs heterostructure.

In summary, we have studied the effect of a unidirectional
periodic potential on the orientation of the UCDW state. By
considering two electric subbands of a wide quantum well, it is
found that the stripe is parallel to the potential for weak
modulation and is perpendicular for strong modulation. For
intermediate modulation, the orientation can be either parallel or
perpendicular to the potential. When the electron density
increases, the stripes experience a rotation of $90^0$ from
parallel or orthogonal to the external potential. The result is
consistent with the recent experimental observation. Our
discussions may help to discern the pinning mechanism of stripes
at half-filled high Landau levels.

\begin{center}
FIGURES
\end{center}

Figure 1 A schematic description of the energy levels with two
electric subbands. $(mn)$'s indicate the two indices of electric
subbands $(m)$ and Landau levels $(n)$. Thin dashed lines are the
corresponding Zeeman splitting. Thick dashed line is the Fermi
level for $\nu=9/2$.

Figure 2 The cohesive energy of the UCDW versus the modulation
strength $V_0/\hbar\Omega$ of the potential. "para" and "orth"
denote parallel and orthogonal orientation to the periodic
potential, respectively. (a) is for $\omega_c/\Omega=0.2941$ and
(b) is for $\omega_c/\Omega=0.8824$.

Figure 3 Anisotropy energy $E_a$ versus $\omega_c/\Omega$ or electron
density (in arbitrary units). $E_a$ is the energy difference between
the parallel orientation and the orthogonal orientation.
$E_a$ is definitely negative for $V_0/\hbar\Omega=0.02$ whereas
definitely positive for $V_0/\hbar\Omega=0.07$. For $V_0/\hbar\Omega=0.05$,
$E_a$ changes from positive to negative as $\omega_c/\Omega$
(or electron density) increases, implying a phase transition from
the orthogonal orientation to the parallel orientation.

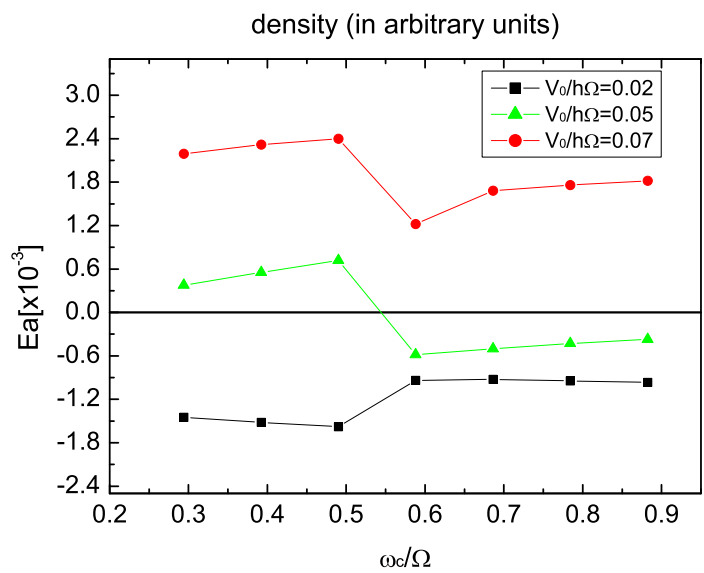
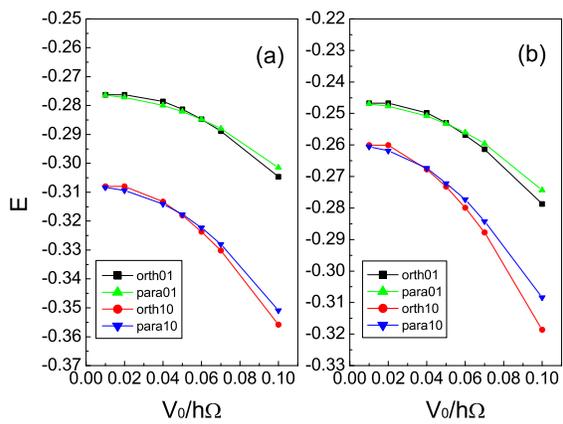
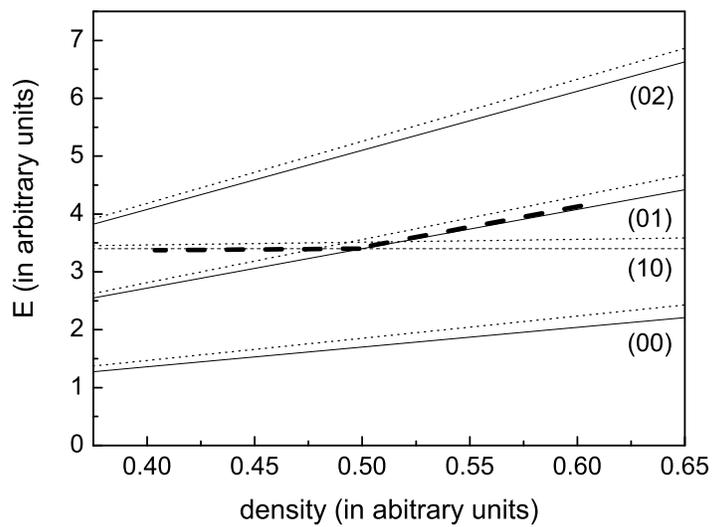
\end{document}